\documentclass[aps,showpacs,preprintnumbers,amsmath,amssymb,floatfix]{revtex4}
\usepackage{graphicx}
\usepackage{latexsym}
\usepackage{verbatim,times,bbm}
\usepackage{color}

\begin{document}

\title{Entanglement enhanced bit rate over multiple uses of a lossy bosonic channel with memory}

\author{C. Lupo and S. Mancini}
\affiliation{Dipartimento di Fisica, Universit\`a di Camerino,
I-62032 Camerino, Italy}

\begin{abstract}

We present a study of the achievable rates for classical information
transmission via a lossy bosonic channel with memory, using homodyne
detection. A comparison with the memoryless case shows that the
presence of memory enhances the bit rate if information is encoded
in collective states, i.e.\ states which are entangled over
different uses of the channel.

\end{abstract}

\pacs{03.67.Hk, 03.65.Yz, 42.50.Dv, 03.67.Mn}

\maketitle

\section{Introduction}\label{s:intro}

One of the tasks of quantum information theory is to evaluate the
capacity of quantum channels for the transmission of classical
information. Recently, much attention has been devoted to the study
of quantum channels with memory (see e.g.\ \cite{werner} and the
references therein). There are evidences that memory effects can
enhance the classical capacity in both the cases of discrete
\cite{discr} and continuous \cite{cont} variables. Bosonic Gaussian
channels provide a fertile benchmark for the study of quantum
channels with continuous alphabet \cite{holevo1,holevo2}. Here we
consider a model of lossy bosonic memory channel introduced in
\cite{oleg}, based on the general scheme proposed in \cite{mancini}
(see also \cite{Ruggeri}). In this kind of model, memory effects
come from the interaction with a common environment containing
correlations. In spite of the fact that each channel belonging to
this family is unitary equivalent to a memoryless one, the presence
of energy constraints can break the unitary symmetry, leaving the
problem of evaluating capacities open (see~\cite{mancini}). A
specific instance of a channel belonging to that family is obtained
by specifying the state of the environment. Here we consider a
multimode squeezed vacuum with one free parameter expressing the
degree of squeezing, this parameter in turn determines the amount of
memory contained in the channel.

The aim of this paper is to evaluate the maximum achievable
transmission rate using homodyne detection. A similar analysis was
already presented in \cite{oleg}, where the maximization was
performed over a specific set of encoding/decoding schemes; here we
optimize the rate over a much larger set. That allows to compare
explicitly the performance of suitably defined {\it collective
schemes}, i.e.\ using entangled states, over {\it local schemes}
involving only separable ones.

The paper proceeds as follows. In section \ref{model} we introduce
the model. In section \ref{ED} we describe the procedure for
encoding and decoding of classical information. The main results are
presented in section \ref{achieve}, where we compare local and
collective encoding/decoding schemes. Section \ref{final} contains
conclusions and comments.

\section{Lossy bosonic channel with memory}\label{model}

\begin{figure}
\centering
\includegraphics[width=0.4\textwidth]{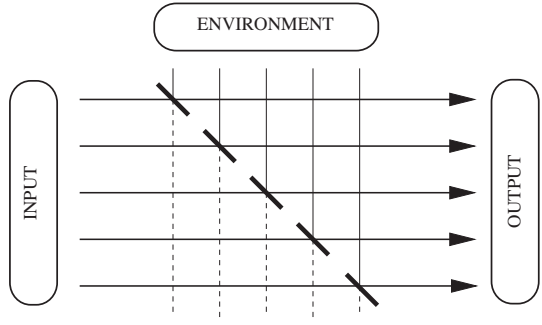}
\caption{A schematic picture of the model of lossy bosonic channel.
Each input mode (left-right line), representing one use of the
channel, interacts with the corresponding environment mode
(top-bottom line) through a beam-splitter. To introduce memory
effects, environment modes are considered in a correlated state.}
\label{fig_model}
\end{figure}

To define our model we need to introduce, for a given integer $n$, a
set of $n$ input bosonic oscillators, with canonical variables $\{
q_k, p_k \}_{k=1, \dots n}$, and a collection of $n$ ancillary
modes, which play the role of the environment, with canonical
variables $\{ Q_k, P_k \}_{k=1, \dots n}$. In the following we refer
to this set of oscillators as the {\em local modes}. All the
frequencies are assumed to be degenerate and normalized to one. The
integer $k$ labels the sequential uses of the channel. At the $k$th
use, the $k$th input mode is linearly mixed with the $k$th
environment mode at a beam splitter with given transmissivity $\eta$
(see figure~\ref{fig_model}). In the Heisenberg picture, the channel
transforms the input field variables as
\begin{eqnarray}
\begin{array}{ccc}
q'_k & = & \sqrt{\eta}\,q_k + \sqrt{1-\eta}\,Q_k \, , \\
p'_k & = & \sqrt{\eta}\,p_k + \sqrt{1-\eta}\,P_k \, .
\end{array}
\end{eqnarray}

Memory effects among different uses of the channel are present if
the corresponding local modes of the environment are correlated.
Here we assume the environment to be in a Gaussian state with zero
mean, described by the Wigner function
\begin{equation}\label{Wigner_res_loc}
W = \frac{1}{\sqrt{\det(V)}} \exp{\left(-\frac{1}{2} X^\mathsf{T}
V^{-1} X \right)},
\end{equation}
where $X^\mathsf{T}:=(Q_1, Q_2, \dots Q_n, P_1, P_2, \dots P_n)$
indicates the vector of coordinates in the environment phase space.
The choice of a Gaussian state for the environment makes the channel
itself Gaussian. We chose a covariance matrix with the following
block-diagonal form
\begin{eqnarray}\label{covm}
V = \frac{1}{2} \left(
\begin{array}{cc}
e^{s\Omega} & \mathbb{O} \\
\mathbb{O}  & e^{-s\Omega}
\end{array} \right).
\end{eqnarray}
This is a {\it bona fide} covariance matrix as long as the matrix
$\Omega$ is symmetric and the parameter $s$ is real. It represents a
multimode squeezed vacuum (see e.g.\ \cite{squeez}). The squeezing
parameter $s$ (or $|s|$), determining how strong environment
correlations are, can be also interpreted as a measure of {\em
memory} between different uses of the bosonic channel. As to the
form of $\Omega$, we chose the following $n \times n$ matrix:
\begin{eqnarray}\label{omega}
\Omega = \left( \begin{array}{cccccc}
0      & 1      & 0      & \dots  & 0      & 0 \\
1      & 0      & 1      & \ddots & 0      & 0 \\
0      & 1      & 0      & \ddots & 0      & 0 \\
\vdots & \vdots & \ddots & \ddots & \vdots & \vdots \\
0      & 0      & 0      & \ddots & 0      & 1 \\
0      & 0      & 0      & \dots  & 1      & 0
\end{array} \right).
\end{eqnarray}
The spectrum of this matrix was already presented in e.g.\
\cite{oleg}, its eigenvalues are
\begin{equation}
\lambda_j = 2 \cos{\left( \frac{\pi j}{n+1}\right)} \ \ \mbox{for} \
\ j=1, \dots n,
\end{equation}
the components of the corresponding eigenvectors being
\begin{equation}
v_{j,k} = \sqrt{\frac{2}{n+1}} \sin{\left(\frac{jk\pi}{n+1}\right)}
\ \ \mbox{for} \ \ k=1,\dots n.
\end{equation}

According to Williamson theorem (see e.g.\ \cite{paris}) one can
always find a set of collective modes, with canonical variables $\{
\tilde{Q}_j, \tilde{P_j} \}_{j=1,\dots n}$, which diagonalize the
covariance matrix. In this basis, the Wigner function has the form
\begin{equation}
\tilde{W} = \frac{1}{\sqrt{\det(\tilde{V})}} \exp{\left(-\frac{1}{2}
\tilde{X}^\mathsf{T} \tilde{V}^{-1} \tilde{X} \right)},
\end{equation}
where $\tilde{X}^\mathsf{T}:=(\tilde{Q}_1, \tilde{P}_1, \dots
\tilde{Q}_n, \tilde{P}_n)$. An explicit expression for these
variables can be obtained from the eigenvectors of the matrix
$\Omega$, yielding to define the following variables:
\begin{eqnarray}\label{collective}
\begin{array}{ccc}
\tilde{Q}_j & := & v_{j,k} Q_k \, , \\
\tilde{P}_j & := & v_{j,k} P_k \, .
\end{array}
\end{eqnarray}
Analogously, we introduce the following collective variables at the
input field:
\begin{eqnarray}\label{point}
\begin{array}{ccc}
\tilde{q}_j & := & v_{j,k} q_k \, , \\
\tilde{p}_j & := & v_{j,k} p_k \, .
\end{array}
\end{eqnarray}
We refer to the set of oscillators with field variables $\{
\tilde{Q}_j, \tilde{P}_j \}_{j=1,\dots n}$ and $\{ \tilde{q}_j,
\tilde{p}_j \}_{j=1,\dots n}$ as the {\em collective modes}. Let us
remark that the action of the channel in the Heisenberg picture on
the collective modes is formally unchanged:
\begin{eqnarray}
\begin{array}{ccc}
\tilde{q}'_j & = & \sqrt{\eta}\, \tilde{q}_j + \sqrt{1-\eta}\, \tilde{Q}_j \, , \\
\tilde{p}'_j & = & \sqrt{\eta}\, \tilde{p}_j + \sqrt{1-\eta}\, \tilde{P}_j \, .
\end{array}
\end{eqnarray}

It is worth noticing that the diagonal covariance matrix $\tilde{V}$
is the direct sum of one-mode covariance matrices:
\begin{equation}
\tilde{V} = \bigoplus_{j=1}^n \tilde{V}_j \, .
\end{equation}
That implies that the state of the environment is simply separable
in the basis of collective modes. It follows from the transformation
(\ref{collective}) that the entries of $\tilde{V}$ are the
eigenvalues of $V$. Using the eigenvalues of the matrix $\Omega$, we
obtain the following expression for the covariance matrix of the
$j$th collective mode:
\begin{eqnarray}
\tilde{V}_j = \frac{1}{2}\left( \begin{array}{cc} e^{s_j} & 0 \\ 0 &
e^{-s_j}
\end{array} \right),
\end{eqnarray}
where $s_j := s \lambda_j$. We observe that each collective
oscillator of the environment is in a squeezed state with $j$
dependent squeezing parameter.

To conclude this section, let us come back to the basis of the local
modes. For a given integer $k_0 \in [1,n]$, by integrating the
Wigner function in (\ref{Wigner_res_loc}) over the local variables
$\{ q_k , p_k \}$ for $k \neq k_0$ we obtain the Wigner function
describing the state of the $k_0$th local mode of the environment.
The corresponding local state is thermal-like, with average number
of excitations:
\begin{equation}
T_\mathrm{eff}(s,k_0) = \frac{1}{2} \left[ \sum_{j=1}^n v_{j,k_0}^2
e^{s_j} \right] - \frac{1}{2} \, .
\end{equation}
By the symmetries of $v_{j,k}$ and $s_j$, this can be rewritten as
follows:
\begin{eqnarray}
T_\mathrm{eff}(s,k_0) = \left\{
\begin{array}{lr}
\left[\sum_{j=1}^{n/2} v_{j,k_0}^2 \cosh{s_j}\right] - \frac{1}{2}\, , & \mbox{if $n$ is even.}\\
\left[\sum_{j=1}^{(n-1)/2} v_{j,k_0}^2 \cosh{s_j}\right] +
\frac{1}{2} v_{(n+1)/2,k_0}^2  - \frac{1}{2}\, , & \mbox{if $n$ is
odd}.
\end{array}
\right.
\end{eqnarray}
Notice that the {\em local temperature} $T_\mathrm{eff}(s,k_0)$ is a
monotonically increasing function of $|s|$.

\section{Continuous variables encoding/decoding}\label{ED}

Classical information is sent via the bosonic channel by choosing a
suitable scheme for encoding (by state preparation of the input
field) and decoding (by observing the output field) a classical
alphabet.

As to the encoding, we introduce a reference pure state of the input
field, described by the Gaussian Wigner function
\begin{equation}\label{inputs}
w = \frac{1}{\sqrt{\det(\sigma)}} \exp{\left(-\frac{1}{2}
x^\mathsf{T} \sigma^{-1} x \right)},
\end{equation}
expressed in terms of the local variable $x:=(q_1, \dots q_n, p_1,
\dots p_n)^\mathsf{T}$. A multivariate Gaussian variable $y :=
(y_{q,1}, \dots y_{q,n}, y_{p,1}, \dots y_{p,n})^\mathsf{T}$, with
probability distribution $P(y)$, zero mean and covariance matrix
$Y$, is encoded in the displaced state of the input field described
by the Wigner function
\begin{equation}
w_y = \frac{1}{\sqrt{\det(\sigma)}} \exp{\left[-\frac{1}{2}
(x-y)^\mathsf{T} \sigma^{-1} (x-y) \right]}.
\end{equation}
Eventually, the state describing the statistical ensemble has the
following Wigner function:
\begin{equation}
\bar{w}_y = \int w_y P(y)\,dy = \frac{1}{\sqrt{\det(\sigma+Y)}}
\exp{\left[-\frac{1}{2} x^\mathsf{T} (\sigma+Y)^{-1} x \right]}.
\end{equation}

To avoid infinite energy we introduce a constraint in the maximum
number of excitations at the input field per channel use (i.e.\ per
mode) in average. Allowing no more than $N$ excitations per mode in
average, the constraint can be written in terms of the covariance
matrices, in natural units:
\begin{equation}\label{energy}
\frac{1}{2n} \mathrm{tr}\left( \sigma + Y \right) \le N +
\frac{1}{2} \, .
\end{equation}

We can summarize the action of the channel on Gaussian states as
follows. For a given value of the displacement amplitudes (hence for
a given letter of the alphabet) the Wigner function at the output
field has covariance matrix
\begin{equation}
\sigma'_y = \eta \sigma + (1-\eta) V.
\end{equation}
On the other hand, the output state averaged over the letters of the
continuous alphabet has covariance matrix
\begin{equation}
\sigma' = \eta ( \sigma + Y ) + (1-\eta) V.
\end{equation}

We can equivalently work in the basis of collective modes, in which
the covariance matrix of the reference input state is denoted as
$\tilde{\sigma}$. Analogously, the zero mean Gaussian variable
$\tilde{y}:=(\tilde{y}_{q,1}, \dots \tilde{y}_{p,n})^\mathsf{T}$,
defined by the relations
\begin{eqnarray}
\begin{array}{ccc}
\tilde{y}_{q,j} & := & v_{j,k} y_{q,k} \, , \\
\tilde{y}_{p,j} & := & v_{j,k} y_{p,k} \, ,
\end{array}
\end{eqnarray}
has covariance matrix $\tilde{Y}$. In the collective basis, the
output field is described by the covariance matrices
\begin{equation}
\tilde{\sigma}'_y = \eta \tilde{\sigma} + (1-\eta) \tilde{V}
\end{equation}
and
\begin{equation}
\tilde{\sigma}' = \eta ( \tilde{\sigma} + \tilde{Y} ) + (1-\eta)
\tilde{V}.
\end{equation}
It is worth noticing that the form of the energy constraint is
preserved, namely
\begin{equation}
\frac{1}{2n} \mathrm{tr}\left( \tilde{\sigma} + \tilde{Y} \right)
\le N + \frac{1}{2} \, .
\end{equation}

Homodyne detection requires the choice of a compatible set of $n$
quadratures to be measured at the output field. A generic set of
quadratures is
\begin{equation}\label{homodyne}
\{ z_h \ \ | \ \ z_h = \mathbb{R}_{hk} q_k + \mathbb{S}_{hk} p_k \}
\end{equation}
for any pair of $n \times n$ matrices $\mathbb{R}$, $\mathbb{S}$
satisfying the relations
$\mathbb{R}\mathbb{R}^\mathsf{T}+\mathbb{S}\mathbb{S}^\mathsf{T}=\mathbb{I}$
and
$\mathbb{R}\mathbb{S}^\mathsf{T}-\mathbb{S}\mathbb{R}^\mathsf{T}=\mathbb{O}$
(see e.g.\ \cite{paris}). Assuming ideal homodyne, the distribution
of the stochastic variable $z:=(z_1, \dots z_n)^\mathsf{T}$ is
Gaussian with zero mean. Its covariance matrix, which we denote as
$Z$, can be computed from the output field covariance matrix using
the relations in~(\ref{homodyne}).

For given covariance matrices $V$, $\sigma$, and for a given set of
quadratures to be measured, a classical channel is defined from the
quantum  one. The capacity of the classical channel can be computed
as the maximum of the mutual information between the output and the
input variables:
\begin{equation}
I(z;y) = H(z) - H(z|y),
\end{equation}
where $H$ denotes the Shannon entropy. The maximization has to be
taken over all possible expressions of the covariance matrix $Y$
compatible with the energy constraint.

\section{Achievable bit rates with homodyne
detection}\label{achieve}

Let us now consider the form of the Wigner function for the
reference input state, determined by the covariance matrix $\sigma$,
and the form of the distribution of the classical variable $y$,
determined by the covariance matrix $Y$. Here we distinguish and
compare two cases: the first one corresponds to a {\em local}
encoding in which $\sigma$ and $Y$ are diagonal; the second one is a
{\em collective} encoding in which $\tilde{\sigma}$ and $\tilde{Y}$
are diagonal. From the view point of the local modes, in the local
encoding scheme information is always carried by simply separable
(unentangled) states, while the collective encoding deals with
states which are in general entangled.

Let us first describe the case of local encoding. In the basis of
the local modes, the diagonal $\sigma$ can be parameterized as
follows
\begin{equation}
\sigma = \bigoplus_{k=1}^n \sigma_k.
\end{equation}
We introduce the real parameters $\{ r_k \}$, such that
\begin{eqnarray}
\sigma_k = \left( \begin{array}{cc} \langle q_k^2 \rangle & \frac{1}{2}\langle q_k p_k + p_k q_k \rangle \\
\frac{1}{2}\langle q_k p_k + p_k q_k \rangle & \langle p_k^2 \rangle
\end{array} \right) =
\frac{1}{2}\left( \begin{array}{cc} e^{r_k} & 0 \\
0 & e^{-r_k} \end{array} \right).
\end{eqnarray}
Analogously, the covariance matrix $Y$ reads
\begin{equation}
Y = \bigoplus_{k=1}^n Y_k
\end{equation}
and is parameterized by the positive parameters $\{ c_{q,k}, c_{p,k}
\}$:
\begin{eqnarray}
Y_k = \frac{1}{2}\left( \begin{array}{cc} c_{q,k} & 0 \\
0 & c_{p,k} \end{array} \right).
\end{eqnarray}
As to the decoding part, the natural choice is to measure the local
quadratures
\begin{equation}
z_k := \cos{\theta_k} q_k + \sin{\theta_k} p_k \, .
\end{equation}

The covariance matrix of the output variable $z:=(z_1,\dots
z_n)^\mathsf{T}$ can be easily computed. It is diagonal as well,
with the variances:
\begin{equation}
\langle z_k^2\rangle = \cos^2{\theta_k} \left[ \eta
\left(\frac{e^{r_k}}{2}+c_{q,k}\right) + (1-\eta)
\left(T_\mathrm{eff}(s,k) + \frac{1}{2}\right) \right] +
\sin^2{\theta_k} \left[ \eta
\left(\frac{e^{-r_k}}{2}+c_{p,k}\right)+ (1-\eta)
\left(T_\mathrm{eff}(s,k) + \frac{1}{2}\right) \right].
\end{equation}
The output variable has conditioned covariance matrix, denoted
$Z_y$, which is diagonal with entries
\begin{equation}
\langle z_k^2\rangle_y = \cos^2{\theta_k} \left[ \eta
\frac{e^{r_k}}{2} + (1-\eta) \left(T_\mathrm{eff}(s,k) +
\frac{1}{2}\right) \right] + \sin^2{\theta_k} \left[ \eta
\frac{e^{-r_k}}{2} + (1-\eta) \left(T_\mathrm{eff}(s,k) +
\frac{1}{2}\right) \right].
\end{equation}

Since all the probability distributions are Gaussian, the mutual
information can be easily written in terms of the covariance
matrices:
\begin{equation}
I(z;y) = \frac{1}{2}\log_2{\left[\det(Z)\right]} - \frac{1}{2}
\log_2{\left[\det(Z_y)\right]} = \frac{1}{2}\sum_{k=1}^n
\log_2{\left(\frac{\langle z_k^2\rangle}{\langle
z_k^2\rangle_y}\right)}.
\end{equation}
The maximization of the mutual information is over the $3n$
parameters $r_k, c_{q,k}, c_{p,k}$ satisfying the constraints
\begin{eqnarray}
\frac{1}{2n} \sum_{k=1}^n \left( \cosh{r_k} + c_{q,k} + c_{p,k}
\right) & = & N_k + \frac{1}{2} \, , \\
\frac{1}{n} \sum_{k=1}^n N_k \le N,
\end{eqnarray}
and over the $n$ angles $\theta_k$, which determine the chosen
quadratures. Concerning the choice of the optimal quadratures, it is
immediate to recognize that the maximum is reached for
$\sin{\theta_k}=0$ if $r_k > 0$, and for $\cos{\theta_k}=0$ if $r_k
< 0$, while the value of $\theta_k$ is irrelevant if $r_k=0$.
Maximizing over the remaining variables and applying the additivity
of mutual information, we obtain the following expression for the
maximum mutual information with local encoding/decoding scheme:
\begin{equation}\label{maxx_loc}
F_\mathrm{loc} = \max_{\{N_k\}} \left\{ \frac{1}{2} \sum_{k=1}^n
\log_2{\left[ 1 + 2
\frac{2N_k+1-\cosh{r_k^\mathrm{opt}}}{e^{-r_k^\mathrm{opt}}+\nu
(2T_\mathrm{eff}(s,k)+1)}\right]} \ \ | \ \ \frac{1}{n}\sum_{k=1}^n
N_k \le N \right\},
\end{equation}
where
\begin{equation}
e^{r_k^\mathrm{opt}} = \frac{1}{\nu(2T_\mathrm{eff}(s,k)+1)} \left[
\sqrt{ 1 + \nu (2T_\mathrm{eff}(s,k)+1) \left( \nu
(2T_\mathrm{eff}(s,k)+1) + 4N_k+2  \right) } - 1 \right],
\end{equation}
and we have defined $\nu := (1-\eta)/\eta$.

Let us now consider the case of collective encoding. In the basis of
the collective modes, the diagonal $\tilde{\sigma}$ can be
parameterized as follows
\begin{equation}
\tilde{\sigma} = \bigoplus_{j=1}^n \tilde{\sigma}_j
\end{equation}
with
\begin{eqnarray}
\tilde{\sigma}_j = \frac{1}{2}\left( \begin{array}{cc} e^{\tilde{r}_j} & 0 \\
0 & e^{-\tilde{r}_j} \end{array} \right).
\end{eqnarray}
Analogously, the covariance matrix $\tilde{Y}$ is
\begin{equation}
\tilde{Y} = \bigoplus_{j=1}^n \tilde{Y}_j
\end{equation}
with
\begin{eqnarray}
\tilde{Y}_j = \frac{1}{2}\left( \begin{array}{cc} \tilde{c}_{q,j} & 0 \\
0 & \tilde{c}_{p,j} \end{array} \right).
\end{eqnarray}

As to the decoding part, the natural choice is to measure the
collective quadratures defined as
\begin{equation}
\tilde{z}_j = \cos{\theta_j} \tilde{q}_j + \sin{\theta_j}
\tilde{p}_j \, .
\end{equation}
The output variable $\tilde{z}:=(\tilde{z}_1, \dots
\tilde{z}_n)^\mathsf{T}$ has diagonal covariance matrix $\tilde{Z}$
and conditional covariance matrix $\tilde{Z}_y$. The corresponding
variances read
\begin{equation}
\langle \tilde{z}_j^2 \rangle = \cos^2{\theta_j} \left[ \eta
\left(\frac{e^{\tilde{r}_j}}{2}+\tilde{c}_{q,j}\right) + (1-\eta)
\frac{e^{s_j}}{2} \right] + \sin^2{\theta_j} \left[ \eta
\left(\frac{e^{-\tilde{r}_j}}{2}+\tilde{c}_{p,j}\right)+ (1-\eta)
\frac{e^{-s_j}}{2} \right]
\end{equation}
and
\begin{equation}
\langle \tilde{z}_j^2 \rangle_y = \cos^2{\theta_j} \left[ \eta
\frac{e^{\tilde{r}_j}}{2} + (1-\eta) \frac{e^{s_j}}{2} \right] +
\sin^2{\theta_j} \left[ \eta \frac{e^{-\tilde{r}_j}}{2} + (1-\eta)
\frac{e^{-s_j}}{2} \right].
\end{equation}

It is easy to recognize that the maximum of the mutual information
is reached for $\sin{\theta_j}=0$ if $s_j>0$, and for
$\cos{\theta_j}=0$ if $s_j<0$, finally if $s_j=0$ we can argue as in
the case of local encoding. By maximizing over the remaining $3n$
variables $\{ \tilde{r}_j$, $\tilde{c}_{q,j}$, $\tilde{c}_{p,j} \}$
under the constraints
\begin{eqnarray}
\frac{1}{2n} \sum_{j=1}^n \left( \cosh{\tilde{r}_j} +
\tilde{c}_{q,j} + \tilde{c}_{p,j} \right) & = & N_j + \frac{1}{2} \, , \\
\frac{1}{n} \sum_{j=1}^n N_j \le N
\end{eqnarray}
and applying the additivity of the mutual information, we obtain the
following expression for the maximum rate of transmission with
encoding/decoding in collective variables:
\begin{equation}\label{maxx_coll}
F_\mathrm{coll} = \max_{\{N_j\}} \left\{ \frac{1}{2} \sum_{j=1}^n
\log_2{\left[ 1 + 2
\frac{2N_j+1-\cosh{\tilde{r}_j^\mathrm{opt}}}{e^{-\tilde{r}_j^\mathrm{opt}}+\nu
e^{-|s_j|}}\right]} \ \ | \ \ \frac{1}{n} \sum_{j=1}^n N_j \le N
\right\}.
\end{equation}
The optimal value $\tilde{r}_j^\mathrm{opt}$ is determined by the
equation
\begin{equation}
e^{\tilde{r}_j^\mathrm{opt}} = \frac{e^{|s_j|}}{\nu} \left[ \sqrt{ 1
+ \nu e^{-|s_j|} \left( \nu e^{-|s_j|} + 4N_j+2  \right) } - 1
\right].
\end{equation}

\begin{figure}
\centering
\includegraphics[width=0.4\textwidth]{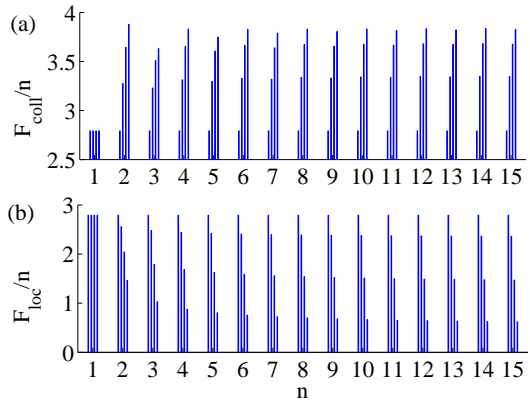}
\caption{The histograms show the maximum achievable rates per
channel use with collective (a) and local (b) encoding/decoding
scheme, as function of the number of channel uses. For each value of
$n$, the histograms correspond to different values of the memory
parameter, from the left to the right $s=0$, $s=1$, $s=2$, $s=3$.
The maximum average excitations per mode is $N=8$, the
transmissivity $\eta=0.7$.} \label{bars}
\end{figure}

\begin{figure}
\centering
\includegraphics[width=0.4\textwidth]{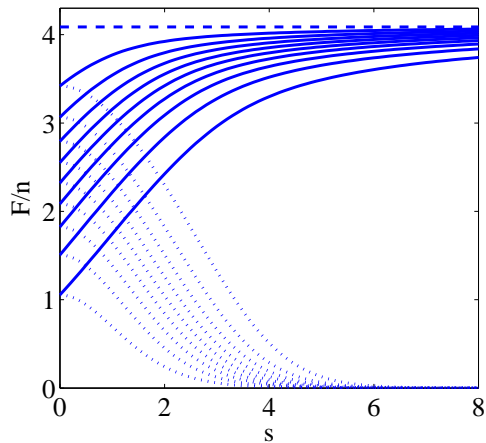}
\caption{The plot shows the achievable rates per channel use with
collective (solid line) and local (dotted line) encoding/decoding
scheme, as function of the memory parameter, for $n=10$ uses of the
channel. The curves correspond to different values of the
transmissivity, from the bottom to the top $\eta$ varies from
$\eta=0.1$ to $\eta=0.9$ by steps of $0.1$. The maximum average
excitations per mode is $N=8$, which corresponds to the asymptotic
value (plotted in dashed line) of the rate with collective scheme
(see equation (\ref{limit})): $\lim_{s\rightarrow\infty}
F_\mathrm{coll}/n = \log_2(2N+1)\simeq 4.0875$.} \label{rates}
\end{figure}

The maximum mutual informations {\em per channel use}, respectively
defined as $F_\mathrm{loc}/n$ and $F_\mathrm{coll}/n$, are plotted
in figure \ref{bars} as function of number of uses of the channel,
for several values of the memory parameter. The case $n=1$ or $s=0$
correspond to the memoryless channel. Notice the different behavior
of the rates as the number of uses (or the value of the memory
parameter) increases. The same quantities are plotted together in
figure \ref{rates}, for $n=10$ uses of the channel, as function of
the memory parameter, for several values of the beam splitter
transmissivity. Notice the different behavior of the rates
corresponding to local (dotted line) and collective (solid line)
encoding/decoding scheme.

The expression of the maximal mutual information allows to make
comparison with the memoryless case (see also the plot in figure
\ref{rates}). First of all, let us recall that the memoryless
channel is recovered for $s=0$; our results show that the maximum
mutual information is a monotonically increasing function of $|s|$
in the case of collective encoding/decoding scheme, while it is
monotonically decreasing in the case of the local one. This is
enough to conclude that, in the presence of memory, collective
encoding/decoding is the optimal choice. Secondly, let us notice
that the two schemes are equivalent in the memoryless case, i.e.\
$s=0$ or $n=1$; that leads to conclude that the presence of memory
enhances the bit rate as long as collective encoding/decoding is
allowed.

\section{Conclusion}\label{final}

We have discussed the model of lossy bosonic Gaussian channel with
memory presented in \cite{oleg}, an instance of a general class of
bosonic memory channel introduced in \cite{mancini}. Our model is
characterized by a single parameter which determines the amount of
memory contained in the channel. By considering the case of homodyne
detection, we have compared two different encoding/decoding schemes:
a {\em local} one in which classical information is encoded and
carried by unentangled (simply separable) states; and a {\em
collective} scheme based on states which are in general entangled
among different uses of the channel. We have computed the maximum
achievable bit rate using both the schemes, assuming energy
constrains at the input field, for an arbitrary number of channel
uses. Our results lead to conclude that, as the memory parameter
increases, the collective scheme becomes more and more efficient
than the local one. It is hence interesting to analyze the
asymptotic limit of the rates for $|s|\rightarrow\infty$,
corresponding to {\em infinite} memory. From the expressions in
(\ref{maxx_loc}) and (\ref{maxx_coll}) we obtain
\begin{equation}
\lim_{s\rightarrow\infty} \frac{F_\mathrm{loc}}{n} = 0
\end{equation}
and
\begin{equation}\label{limit}
\lim_{s\rightarrow\infty} \frac{F_\mathrm{coll}}{n} =
\log_2{(2N+1)}.
\end{equation}
Notice that the latter expression holds independently of $n$ and for
any value of $\eta\neq 0,1$. Interestingly enough, the asymptotic
value in (\ref{limit}) equals the maximum achievable rate of the
(memoryless) noiseless channel ($\eta=1$) with homodyne detection
(see \cite{holevo1}). In this sense, by using local
encoding/decoding one sees infinite noise in the channel with high
degree of memory, while by using a collective encoding/decoding one
can completely avoid it.

\acknowledgments C.L. thanks Oleg V. Pilyavets for a critical
reading of the manuscript.

\end{document}